\documentclass[sigconf]{acmart}
\usepackage{graphicx} 
\usepackage{colortbl} 
\usepackage[table]{xcolor} 
\usepackage{booktabs} 
\usepackage[capitalise]{cleveref}

\usepackage{algorithmic}
\usepackage{url}
\usepackage{textcomp}
\usepackage{longtable}
\usepackage{xltabular}
\usepackage{threeparttable}
\usepackage[most]{tcolorbox}

\tcbset{
  interviewquote/.style={
    enhanced,
    breakable,
    colback=gray!5,
    colframe=gray!20,
    left=.2em,
    right=.2em,
    top=0.2em,
    bottom=0.2em,
    boxrule=1pt
  }
}

\newcommand{\interviewquote}[2]{%
  \begin{tcolorbox}[interviewquote]
    \emph{``#1''} #2
  \end{tcolorbox}
}

\tcbset{
  redditquote/.style={
    enhanced,
    breakable,
    colback=blue!5,
    colframe=blue!20,
    left=.2em,
    right=.2em,
    top=0.2em,
    bottom=0.2em,
    boxrule=1pt
  }
}

\newcommand{\redditquote}[2]{%
  \begin{tcolorbox}[redditquote]
    \emph{``#1''} #2
  \end{tcolorbox}
}

\tcbset{
  blogquote/.style={
    enhanced,
    breakable,
    colback=green!5,
    colframe=green!20,
    left=.2em,
    right=.2em,
    top=0.2em,
    bottom=0.2em,
    boxrule=1pt
  }
}

\newcommand{\blogquote}[2]{%
  \begin{tcolorbox}[blogquote]
    \emph{``#1''} #2
  \end{tcolorbox}
}

\setcopyright{rightsretained}
\copyrightyear{2025}
\acmYear{2025}
\acmDOI{XXXXXXX.XXXXXXX}

\begin{document}

\title{An Empirical Investigation of the Experiences of Dyslexic Software Engineers}

\author{Marcos Vinicius Cruz}
\affiliation{%
\institution{Reykjavik University}
  \city{Reykjavik}
  \country{Iceland}}
  \email{marcosc@ru.is}

\author{Pragya Verma}
\affiliation{%
\institution{Reykjavik University}
  \city{Reykjavik}
  \country{Iceland}}
\email{pragyav@ru.is}

\author{Grischa Liebel}
\affiliation{%
\institution{Reykjavik University}
  \city{Reykjavik}
  \country{Iceland}}
\email{grischal@ru.is}


\begin{abstract}
Dyslexia is a common learning disorder that primarily impairs an individual's reading and writing abilities. In adults, dyslexia can affect both professional and personal lives, often leading to mental challenges and difficulties acquiring and keeping work. In Software Engineering (SE), reading and writing difficulties appear to pose substantial challenges for core tasks such as programming. However, initial studies indicate that these challenges may not significantly affect their performance compared to non-dyslexic colleagues. Conversely, strengths associated with dyslexia could be particularly valuable in areas like programming and design. 
However, there is currently no work that explores the experiences of dyslexic software engineers, and puts their strengths into relation with their difficulties.
To address this, we present a qualitative study of the experiences of dyslexic individuals in SE. We followed the basic stage of the Socio-Technical Grounded Theory method and base our findings on data collected through 10 interviews with dyslexic software engineers, 3 blog posts and 153 posts on the social media platform Reddit.
We find that dyslexic software engineers especially struggle at the programming learning stage, but can succeed and indeed excel at many SE tasks once they master this step. Common SE-specific support tools, such as code completion and linters are especially useful to these individuals and mitigate many of the experienced difficulties. Finally, dyslexic software engineers exhibit strengths in areas such as visual thinking and creativity. 
Our findings have implications to SE practice and motivate several areas of future research in SE, such as investigating what makes code less/more understandable to dyslexic individuals.
\end{abstract}



\maketitle

\section{Introduction}
Neurodiversity refers to natural variations in human cognitive, including conditions such as Autism Spectrum Disorder (ASD), Attention Deficit Hyperactivity Disorder (ADHD), and dyslexia. The neurodiversity movement advocates for not treating these variations as deficits that require treatment, but as a source of distinctive cognitive profiles that have their dedicated strengths and perspectives \cite{austin2017neurodiversity}. 

As one of the conditions included in the neurodiversity umbrella term, dyslexia 
is defined as a learning disorder that primarily impairs an individual's ability to recognize words and decode written patterns \cite{dsm-5}. These difficulties result in deficits in both oral and written language skills, including reading and writing speed, and are independent of age, formal education, or socio-cultural background, affecting approximately 7\% of the world’s population \cite{wajuihian2011dyslexia, peterson2012developmental}. 

In adults, dyslexia can affect both professional and personal lives, often leading to low self-esteem \cite{Snowling2020DefiningAU}, emotional challenges and frustrations in their professional and educational pursuits \cite{Reiss2004DevelopmentalDI}, as well as anxiety or emotional disorders \cite{livingston2018developmental}.
It frequently has a greater impact on individuals without a diagnosis, ultimately affecting their confidence and professional prospects. 

In the field of Software Engineering (SE), reading and writing difficulties may appear to pose challenges for core tasks such as programming.
However, initial eye-tracking studies indicate that these challenges may not significantly affect their performance compared to their non-dyslexic counterparts \cite{mcchesney18,mcchesney21b}, potentially due to the constrained and highly structured syntax of programming languages. 
Conversely, strengths associated with dyslexia, such as strong visual thinking and creative problem-solving abilities, could be particularly valuable in areas like programming and design \cite{Gilger2016Reading}. 
However, there is currently no work that explores the experiences of dyslexic software engineers, and puts their strengths into relation with their difficulties.

To address this, we present a qualitative study of the experiences of individuals with dyslexia in SE. We aim to answer the following research question:
\textbf{RQ: What are the experiences of dyslexic individuals in Software Engineering?}
To answer this question, we followed the basic stage of the Socio-Technical Grounded Theory (STGT) method \cite{hoda2021socio}. Based on data collected through 10 interviews with dyslexic software engineers, 3 blog posts and 153 posts on the social media platform Reddit\footnote{https://www.reddit.com} and their corresponding comments, we find that dyslexic software engineers face difficulties in reading and writing, as expected.
This affects their ability to learn programming and some programming-related tasks, such as reading other people's code.
However, there are also a large amount of voices that demonstrate that mastering programming and building a successful SE career is possible.
To address the inherent reading and writing challenges, individuals use various generic support tools, such as spell-checkers, text-to-speech tools or generative AI support. Additionally, they use strategies in their workplace, such as mentors and taking notes.
Additionally, various support tools and strategies are used that specifically relate to SE, in particular programming-related tools such as code completion, linters, compilers and customized IDEs.
Similarly, individuals mention that programming-related strategies are a substantial help, such as using statically-typed languages, thinking through problems using diagrams or pseudo code, and using specific naming conventions for variables and functions.
Finally, individuals report various cognitive strengths that enable them to excel in SE-related tasks, such as divergent thinking and visual thinking capabilities.

\section{Background and related work}
\label{sec:rw}
This section provides a overview of the existing literature on dyslexia and how it relates to SE.

\subsection{Dyslexia}
\label{sec:background}
Dyslexia is a learning disability that persistently affects an individual’s ability to spell and decode written texts \cite{dsm-5}. 
Dyslexia can occur across the full range of intellectual abilities \cite{Snowling2020DefiningAU}.
The predominant theory is the phonological deficit hypothesis, positing that dyslexia stems from difficulties in processing the sound structure of language \cite{articleS, Reiss2004DevelopmentalDI, Snowling2020DefiningAU}. This deficit affects the ability to map letters to sounds, which is an elementary skill for reading \cite{articleS}.

Although dyslexia can be managed through various strategies, it continues to pose significant challenges \cite{articleS}.  
Dyslexia can also manifest in problems with verbal short-term memory, word retrieval, and with learning new spoken words \cite{Snowling2020DefiningAU}.
Dyslexia is frequently co-morbid (i.e., it frequently co-occurs) with other neurodevelopmental conditions, such as ADHD and ASD \cite{dsm-5}.

\subsection{Dyslexia in the Workplace}

The neurodiversity movement encourages viewing neurodivergent individuals in the light of cognitive diversity. As a consequence, this perspective means that these individuals also possess unique strengths rather than just challenges \cite{LeFevreLevy2023NeurodiversityIT}. As such, many skills associated with dyslexia are becoming increasingly valuable as routine tasks are automated, such as creativity, innovative thinking, and seeing the ``bigger picture''\cite{LeFevreLevy2023NeurodiversityIT,powell2004dyslexia}. For example, the UK intelligence agency GCHQ actively recruits individuals with dyslexia for their ability to ``cut through complexity and find original ways to solve problems'' \cite{LeFevreLevy2023NeurodiversityIT}.  
Despite potential advantages, adults with dyslexia face high rates of unemployment and underemployment, making workplace inclusion and support essential \cite{Doyle342584}.

There is limited work that highlights the experiences of dyslexic individuals at workplace and the coping strategies they use to navigate the difficulties faced. In their review, Beer et al. identify various factors that can affect the work participation of individuals with dyslexia. Some of these reported factors are mental functions, including memory and emotional functions, higher cognitive level functions; learning and applying knowledge, including reading, writing; communication, including speaking and writing messages \cite{de2022factors}. To understand the experiences of dyslexic teachers and the coping strategies they use to overcome various challenges, Burns et al. conducted narrative interviews with tertiary teachers with dyslexia \cite{burns2013resilience}. Their findings suggest that teachers with dyslexia struggle with reading and writing difficulties while preparing for presentations and sessions and reading assignments. To overcome these difficulties, they often use visualization techniques, seek support from their social network at work, and nurture their self-esteem and efficacy. Newlands et al. conducted a qualitative study on the experiences of doctors with dyslexia \cite{newlands2015foundation}. Their results suggest that doctors with dyslexia often struggle with different forms of communication involving reading and writing, time management, and anxiety. The authors also identified several coping strategies used by doctors with dyslexia. Some of these strategies include creating lists, using search engines to verify spellings, task prioritization, and planning.  Finally, Major et al. study the experiences of nurses with dyslexia through 14 in-depth semi-structured interviews \cite{major2019effects}.  The findings suggest that nurses with dyslexia face difficulties in documentation, proof-reading their work, and reading in general. To navigate these challenges, nurses with dyslexia often try to avoid words that they find difficult to spell, spend time going over important information, and read documents prior to any meeting. 

Research on supporting dyslexic adults in professional settings is limited, with a historical focus on children and education \cite{Doyle342584,Reiss2004DevelopmentalDI}. One key area of workplace support involves accommodations such as coaching, designed to help employees manage challenges related to working memory, organization, and stress  \cite{Doyle342584}. However, Doyle and McDowall \cite{Doyle342584} highlight that there is a significant gap in the evidence regarding the effectiveness of such interventions for adults in the workplace.

\subsection{Dyslexia in SE}
There is only little research that relates specifically to dyslexia and SE.
In the context of programming, McChesney and Bond conducted a series of eye-tracking studies and controlled experiments \cite{mcchesney18,mcchesney2018b,mcchesney20,mcchesney21,mcchesney21b}.
They find that dyslexic individuals do not seem to be affected in the same way when reading code as when reading text, i.e., they do not exhibit over-proportional deficiencies compared to non-dyslexic individuals.
An explanation might be that reading code is significantly different from regular text, e.g., due to indentation, spacing, or the use of limited keywords.

In addition to the work by McChesney and Bond, there exist several studies that interpret potential challenges of dyslexic programmers or dyslexic individuals learning programming based on more general literature or based on experience, i.e., \cite{powell2004dyslexia,González2017Combining,fuertes2016characterization}.
These studies relate common symptoms of dyslexia to programming, e.g., linking deficits in short-term memory to difficulties in remembering code details or variable names; linking spelling difficulties to syntax errors; and linking reading difficulties to impairments in following code flow.


Finally, there is work on building and evaluating specific tools for dyslexic engineers, such as combining visual and textual programming languages \cite{González2017Combining}.

\subsection{Neurodiversity in SE}
In addition to work specific to dyslexia in SE, there are several studies on other conditions included under the neurodiversity term.
Morris et al. study challenges of neurodivergent software engineers~\cite{morris15} through 10 interviews and a follow-up survey, mainly with ASD.
The findings suggest various issues related to work and interpersonal communication.
Gama et al.~\cite{gama2023understanding,liebel2023challenges,gama25_gt_asd_adhd} conducted several qualitative studies on ADHD and ASD in SE.
These studies result in a theory of challenges and strengths of software engineers with ADHD and/or ASD, finding that these individuals struggle with several important SE-related activities, e.g., task organization and estimation, as well as with physical and mental health.
In terms of strengths, they exhibit, e.g., increased creative skills and perform well when solving puzzles.
Finally, the analysis of Reddit posts by Newman et al.~\cite{newman2025get} complements this work with additional challenges and strengths of programmers with ADHD, as well as a survey that provides estimates of how common these are and how they compare to the challenges/strengths of programmers without ADHD.

In summary, while challenges of dyslexic individuals are well-known, there are gaps in the evidence of the experiences of dyslexic individuals in the workplace, as well as the success of workplace interventions targeted towards these individuals.
Research on adult dyslexia remains especially scarce regarding workplace performance and support \cite{Doyle342584,Reiss2004DevelopmentalDI}.
Finally, in SE, existing work on dyslexia focuses exclusively on programming and learning programming.
There is a lack of research investigating the experiences, challenges, and in particular the strengths of dyslexic software engineers. 
This study aims to address this gap by providing evidence from the daily experiences of dyslexic software engineers.

\section{Methodology}
We conducted the basic stage of the Socio-technical Grounded Theory (STGT)  \cite{hoda2021socio} method. As we investigate the experiences of individuals in technical roles, STGT is an appropriate method choice.

\subsection{Study Design}
Many activities in SE are inherently socio-technical in nature, such as many agile practices or daily work in software engineering teams. Additionally, some aspects of dyslexia, such as difficulties in reading documentation or learning challenges, may affect both the technical and social aspects of SE. 
As such, we decided to apply STGT based on rich accounts of dyslexic professionals.
In the study reported here, we only performed the first stage of STGT, i.e., the basic stage.
We obtained an ethics review from the local ethics board under number \anon{SHV2024-066}. 
Our interview guide, informed consent form and the results from the analysis of Reddit posts are shared in the accompanying dataset\footnote{\url{https://zenodo.org/records/17220970}}.

\subsection{Data Collection and Analysis}

\begin{figure*}
    \centering
    \includegraphics[width=\textwidth]{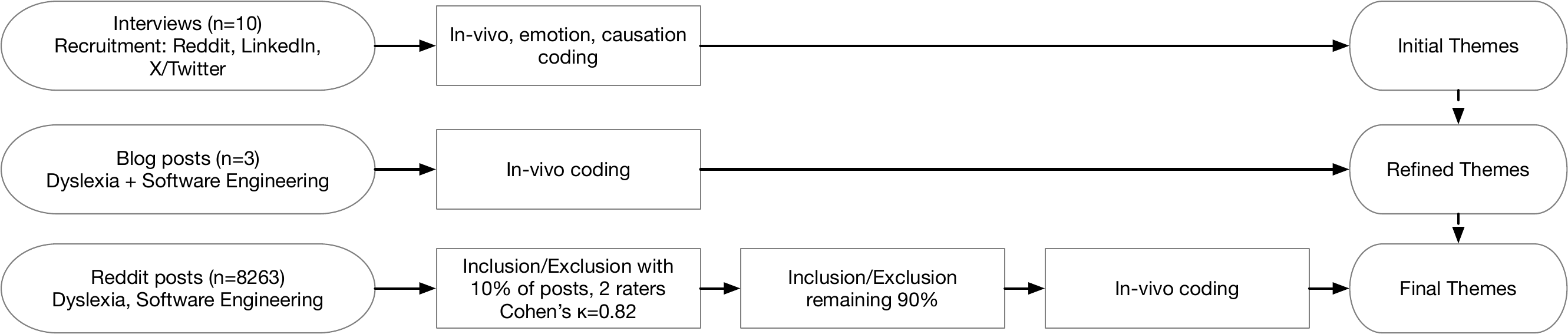}
    \caption{Data Collection and Analysis Steps. The rounded rectangles on the left depict data sources. The rectangles in the middle depict analysis steps. Finally, the rounded rectangles on the right depict the resulting themes.}
    \Description{We used interviews, blog posts and reddit posts and comments as data sources. Initially, 10 interviews were conducted and coded using in-vivo, emotion, and causation coding. Second, we identified 3 online blogs on dyslexia and SE and analysed these using in-vivo coding, resulting in refined themes. Finally, we collected 8263 Reddit posts from dyslexia-related and SE-related subreddits, performed an inter-rater agreement step resulting in a Cohen's Kappa of 0.82, then in-vivo coded all relevant posts and comments, resulting in the final themes.}
    \label{fig:method}
\end{figure*}

Figure~\ref{fig:method} depicts the overall data collection and analysis steps.

\paragraph{Interviews:} We initially collected data through interviews only.
To do so, we created an interview instrument covering various questions that we expected to arise in the context of SE and dyslexia, e.g., communication, productivity, and environmental factors.
The instrument was refined through a pilot with a neurodivergent specialist in SE who participated in a 50-minute online interview. 
We invited interviewees through social media platforms, i.e., LinkedIn, Reddit and X/Twitter. Individuals signed a consent form that was provided online, and included dyslexic-friendly font and style, as well as the option to read the text out loud. Interviewees were guaranteed full anonymity, with the right to withdraw from the study at any time. No compensation was provided. All interviews were conducted remotely,  recorded and transcribed with the help of electronic tools. These transcripts were then cleaned and corrected manually.
Finally, we conducted 10 interviews, whose profiles are listed in Table~\ref{tab:interviewees}. 
Our interviewees' ages ranged from 17 to 53 years, 7 men and 3 women.
They were from three continents and their educational level varied considerably, from elementary school to PhD. The interviewees' roles included (senior) software engineers, professors (with an industrial background), a manager, a student and one intern.

\begin{table}[ht]
\centering
\caption{Interviewee Profiles}
\label{tab:interviewees}
\resizebox{250pt}{!}{
\begin{tabular}{c c c c}
\toprule
\textbf{Interviewee} & \textbf{Age} & \textbf{Educational Level} & \textbf{Role / Area of Experience} \\
\midrule
I1  & 30-39 & Bachelor’s Degree               & Senior Software Engineer \\
I2  & 20-29 & Master’s Degree                 & Student \\
I3  & 50-59 & Master's Degree                             & Professor \\
I4  & 40-49 & PhD                             & Professor \\
I5  & 40-49 & MBA                             & Manager \\
I6  & 30-39 & Bachelor’s Degree               & Senior Software Engineer \\
I7  & 20-29 & Bachelor’s Degree               & Software Engineer \\
I8  & 30-39 & High School                     & Senior Software Engineer \\
I9  & 30-39 & Elementary School               & Senior Software Engineer \\
I10 & <18 & Associate Degree & Intern \\
\bottomrule
\end{tabular}
}
\end{table}

We analyzed the interviews using a mix of in-vivo, emotion and causation coding in the qualitative analysis program QualCoder\footnote{https://qualcoder.wordpress.com}.
In-vivo coding uses the interviewees' own words as codes, while emotion coding highlights specifically emotions voiced by interviewees and causation coding highlights cause-effect statements \cite{saldana2021coding}.
We used emotion coding and causation coding in addition to in-vivo coding to explicitly incorporate emotions (as neurodivergent individuals often struggle with mental health issues) and to encode when individuals attribute issues and strengths directly to dyslexia.
We then organized and discussed resulting codes into basic themes using a whiteboard software, with all three authors discussing these themes.

\paragraph{Blogs and Reddit Data:} Since we only succeeded in recruiting 10 interviewees, and since no did not reach saturation in the themes, we decided to complement the interview data with further sources.
First, we used internet search to identify blogs in which dyslexic software engineers shared their experiences, resulting in 3 blogs \cite{blog1,blog2,blog3}.
We then in-vivo coded the blog posts and added relevant codes to the initial themes.
Second, we extracted relevant posts and associated comments from the social media platform Reddit\footnote{\url{https://reddit.com}}.

Specifically, through manual search we identified sub-reddits relatex to dyslexia (\textit{r/Dyslexia}, \textit{r/DyslexicParents}), neurodiversity (\textit{r/Neurodiversity}, \textit{r/LearningDisabilities}), programming (\textit{r/learnprogramming}, \textit{r/programming}, \textit{r/CodeOpinion}, \textit{r/webdev}, \textit{r/gamedev}, \textit{r/coding}, \textit{r/AskProgramming}), computer science (\textit{r/computerscience}, \textit{r/cscareerquestions}, \textit{r/compsci}) and SE (\textit{r/SoftwareEngineering}, \textit{r/softwareengineer}, \textit{r/Everything\_QA}, \textit{r/QualityAssurance}, \textit{r/softwaretesting}, \textit{r/EngineeringManagers}, \textit{r/ExperiencedDevs}, \textit{r/softwarearchitecture}, \textit{r/softwaredevelopment}, \textit{r/QualityEngineering}, \textit{r/agile}, \textit{r/software\_design}).
In sub-reddits related to SE, computer science and programming, we searched for keywords related to dyslexia.
In neurodiversity sub-reddits we searched for keywords related to dyslexia and SE tasks.
Finally, in dyslexia sub-reddits we searched for keywords related to SE tasks only.

This search resulted in 527 posts and comments from neurodiversity sub-reddits, 2747 from dyslexia sub-reddits, 943 from computer science sub-reddits, 3962 from programming sub-reddits and 84 from SE sub-reddits.
As these data included many irrelevant posts, we developed an inclusion/exclusion protocol similar to those used in systematic literature reviews.
To be included, posts had to be authored by dyslexic individuals or clearly include actual experiences of dyslexic individuals, and contain experiences, challenges, or perspectives related to SE. Posts written by dyslexic authors that focused purely on technical problem-solving without reference to SE were excluded.
Posts authored by non-dyslexic individuals were included when they had the potential to generate relevant responses from dyslexic individuals, e.g., when asking for the experiences of those individuals. Posts with an educational content were also included, if they had the potential to contain relevant insights to the SE profession.
We selected 10\% of all posts in each of the sub-reddit categories and two raters decided independently whether they would be included in the final coding stage, resulting in a Cohen's kappa of $0.82$, which we deemed acceptable.
One author then proceeded to decide which of the remaining posts to include/exclude, and to code the included posts as well as comments to those posts using in-vivo coding.
The coded posts/comments were iteratively incorporated into the existing themes, and the themes refined.
The remaining two authors were involved in this step, discussing and refining themes iteratively.
While adding these posts, the themes saturated, in the sense that no additional sub-themes were added.
No central category emerged during this process, which is why we decided not to advance to the second stage of STGT.
Due to the large amount of different themes, we report in this paper only those related to SE-specific tasks, e.g., programming or debugging.
That is, we exclude general workplace experiences that were reported.

\subsection{Validity Threats}
The data we collected and analyzed relies on subjective accounts of dyslexic individuals, and might thus be biased.
These might affect especially perceived strengths and challenges in relations to others.
As such, follow-up research should complement our work with other forms of evidence, e.g., by obtaining feedback on the workplace performance of dyslexic software engineers from colleagues or managers.
Additionally, future comparative studies might be required to better understand to what extent the reported strengths and challenges compare to non-dyslexic individuals.

Our analysis contains interpretations by the involved researchers, and is therefore prone to researcher bias.
To mitigate this effect to some extent, we tested and followed a strict inclusion/exclusion protocol for the Reddit posts and provide the full search query and dataset online\footnote{\url{https://zenodo.org/records/17220970}}.
This included to only consider posts that are based on actual experience.
Additionally, we excluded comments or posts that we judged to have a strong likelihood of being AI-generated.

As discussed in Section~\ref{sec:background}, dyslexia is often co-morbid with other disorders such as ADHD and ASD. As such, reported experiences might stem from symptoms associated with these other disorders.
Unless this was clearly stated by individuals, we nevertheless include these experiences in our findings, as they are a part of the natural cognitive diversity.

As is common with qualitative studies, we make no claims with respect to the external validity of our findings.
Instead, the value of this study lies in the rich real-world context that we report.

\section{Results}

\begin{figure*}
    \centering
    \includegraphics[width=\textwidth]{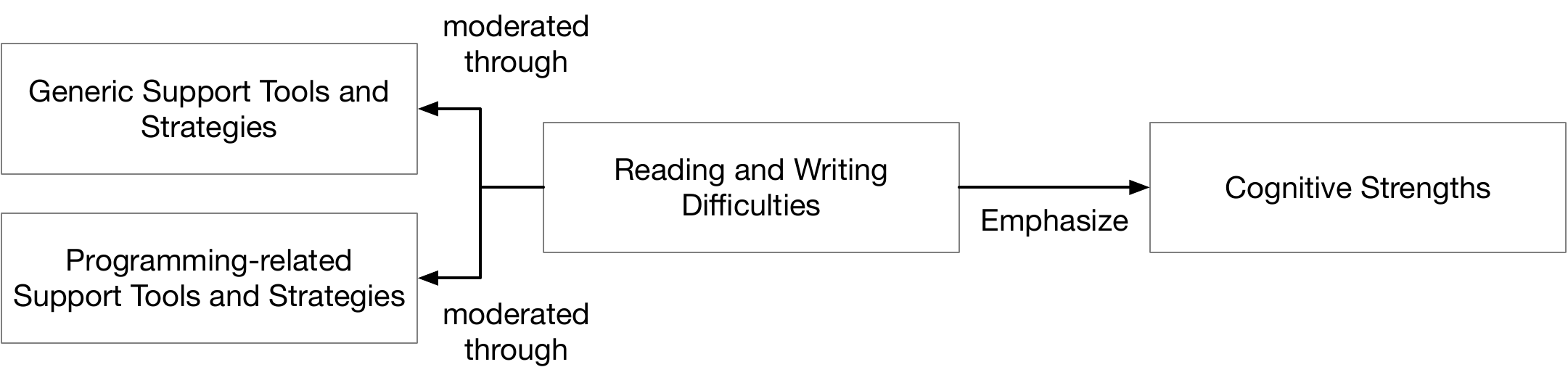}
    \caption{Initial Categories on Experiences of Dyslexic Software Engineers.}
    \Description{The central category describes reading and writing difficulties, including those related to programming and learning how to program. These difficulties are moderated by the use of various generic and programming-related support tools and strategies used by dyslexic engineers. Finally, difficulties in reading and writing emphasize instead various cognitive strengths that are valuable in software engineering.}
    \label{fig:mapping}
\end{figure*}

Our initial theory consists of four high-level categories that influence each other, as depicted in Figure~\ref{fig:mapping}.
The central category, \emph{Reading and Writing Difficulties}, describes issues arising due to well-known dyslexia symptoms.
This category also describes how our sources described their experiences in (learning) programming and programming-related tasks.
To address their difficulties, our sources use various \emph{Generic Support Tools and Strategies}, such as using text-to-speech software.
However, they additionally use various \emph{Programming-related Support Tools and Strategies}, such as employing syntax completion or linters. 
Finally, and potentially due to their difficulties with reading and writing, our sources report on various \emph{Cognitive Strengths} that are highly valuable in SE.

%

In the following, we describe these four categories as well as their sub-categories in detail.

\subsection{Reading and Writing Difficulties}
Our findings show that dyslexic SE professionals face various challenges in the workplace which we discuss next. 


Many dyslexic professionals on Reddit and our interviewees identified themselves as slow readers and learners:
\interviewquote{I'm really a slow reader, and when I need to read articles and stuff, I always need to read the article like 2 times to really get it.}{~- I2}
\interviewquote{I read a little slower than most people.}{~- I8}
In relation to SE, individuals mention this difficulty especially related to reading documentation:
\interviewquote{Oh no, I need to go through 30 pages of reading just to figure out what I need to do [...] I fear documentation the most.}{~- I2}
\redditquote{[...] reading slows me down [...] my main problem is that everything I need to learn is in documentation.}{~- DYS023}
\blogquote{Dyslexia typically involves difficulties with reading, which can slow down the absorption of written information, including documentation and code.}{~- Blog 2}


Naturally, this affects at least to some extent their programming abilities and the process of learning to program:
\redditquote{My main feedback is I’m a really slow programmer.}{~- DYS013}
\redditquote{[Initially,] I couldn't program my way out of a wet paper bag}{~- DYS040}
\redditquote{was having really hard time working with code and cli tools}{~- DYS053}
%

Interestingly, several individuals comment specifically on the learning process, decoupled from programming itself:
\redditquote{I usually just read a small piece of text, and then practice the lines from the book. The programming itself isn't as hard as learning about them [...]}{~- PROG005}
%

One individual commented that the difficulties did not relate to programming per se, but rather learning in general:
\redditquote{learning to code with dyslexia is just like anything with dyslexia. You just have to understand how you learn}{~- DYS083}

In addition to learning programming, some sources mention difficulties in specific coding-related tasks. For instance, several posts mention difficulties in reading other people's or fixing problems in code:
\redditquote{Understanding code that I haven't written I find very hard [...]}{~- DYS100}
\redditquote{I find it much harder to make sense of someone else's code [...]}{~- PROG038}
%
\redditquote{have both ADHD and Dyslexia, which doesn't necessarily make writing hard, but it makes it exceedingly hard to find and fix problems in code or to find the right code}{~- PROG044}
%

Despite the difficulties, many sources share success stories of how they eventually mastered programming or successfully got a career in SE:
\redditquote{When I started programming it was next to impossible for me to wrap my head around, but I just kept working hard and dumping hours upon hours into my craft}{~- CS038}
\redditquote{I failed 3rd grade, 7th grade dropped out of school because of severe dyslexia ended up going to a coding boot camp for 4 months to learn asp.net full stack development. That was 3 years ago now I'm a lead engineer writing code in several languages}{~- PROG038}
%

Finally, some sources state that they believe dyslexia had no influence or even helped them in picking up SE-related tasks:
\interviewquote{[I] don't find it like holding me back. and as there is a tool for everything nowadays}{~- I7}
 %
\redditquote{I've always felt like my dyslexia or at least some of the ways its shaped me were part of what suited me so well to programming}{~- DYS027}
%
\redditquote{Something about being bad at school things made me assume I would be bad at programming too. Definitely not the case. Which is fortunate, since that is now my career.}{~- DYS027}

\subsection{Generic Support Tools and Strategies}
Our findings suggest that dyslexic SE professionals use various generic support tools and strategies to navigate the challenges they face in the workplace. Here, we report these commonly used support tools and strategies.

Our interviewees state that they frequently use Grammarly to overcome writing difficulties: 
\interviewquote{I have Grammarly, which helps with writing in English.}{~- I3}

To navigate reading difficulties, some sources mention that they prefer to use text-to-speech (TTS) software:
\interviewquote{ I started to notice that reading with Microsoft's screen reader has greatly improved my productivity.}{~- I3}
Many dyslexic SE professionals on Reddit and interviewees mention that they use GenAI tools to overcome reading difficulties and to understand long text: 
\interviewquote{especially nowadays with ChatGpt. I can take a paragraph, or whatever I'm trying to write, and I can run it through there. Ask it to highlight the bullet points.}{~- I6}
\interviewquote{ AI tools really helped me as someone with dyslexia for the reading part. and most of the documentation .}{~- I7}

To navigate the reading and writing difficulties, some sources mention that they prefer to keep things short, which is possible with the use of electronic communication platforms such as Slack:
\interviewquote{I like how slack works, because if it is written, if it's in a conversational form and real time, it kind of encourages people to keep things short.}{~- I8}



In relation to organizing their work, individuals mention that they prefer taking notes:
\interviewquote{I found out that notebooks are my best friend.}{~- I2}
\redditquote{write a lot of notes. Notes will help you organize.}{~- DYS092}




Some sources mention that they speak to their mentors which helps them overcome the difficulties they face at workplace:
\redditquote{[...] perhaps suggest to your mentor that you're more a verbal person [...]}{~- CS033}


Finally, some individuals state that maintaining a positive work environment helps them navigate the difficulties they face:
\interviewquote{being grateful and positive makes people take you more seriously.}{~- I9}
\interviewquote{at work, make jokes. People feel better and less worried.}{~- I10}

\subsection{Programming-related Support Tools and Strategies}
To overcome programming-related difficulties, dyslexic SE professionals use various tools and strategies, which we summarize next.

Many dyslexic professionals on Reddit mention that representing the code pictorially helps them understand it better:
\redditquote{[...] understand the code better when it is pictographically displayed [...]}{~- PROG019}
\redditquote{[...] get a visual layout of code that reflects what it does [..]}{~- DYS098}
\redditquote{working out algorithms and such has been drawing out psuedocode. A lot of the time words just can’t describe what is happening well enough, so by making little 'code clouds,' it helps with the memory aspect of working these algorithms out}{~- DYS094}

Some sources mention that the syntax highlighting feature of IDEs helps them navigate the challenges they face while coding:
\interviewquote{I use text editors that are syntax highlighted [...] it highlights, certain keywords and that helps me to visually, process the code.}{~- I6}
\redditquote{Auto formatting and syntax highlighting are your friends}{~- DYS098} 



In addition to syntax highlighting, some sources mention that they find the spell checker feature or linters in IDEs to be helpful:  
\redditquote{In coding I missspell 'length()' about four times a day, and the compiler finds each time - so no real problem.}{~- DYS090}
\redditquote{any decent IDE will provide a basic level of linting, they will tell you when you have mistyped variable names, missing returns, etc.}{~- PROG036}
\redditquote{The real-time compiler also is good if you tend to focus on the big picture rather then little details}{~- CS015} 

Further, some of the dyslexic SE professionals on Reddit mention that code completion features of IDEs often help them: 
\redditquote{If it [IDE] can do auto complete and show you a list of functions for a class or method those can be big helps.}{~- DYS022}
\redditquote{If you use a ide that auto-populates for you it's even better.}{~- DYS086}

We also found that some of the dyslexic professionals on Reddit and our interviewees find debuggers to be helpful:
\redditquote{What really helped me was using the debugger,}{~- DYS029}
\interviewquote{I like the most about code was debugging,}{~- I2}



In relation to programming efficiently, individuals mention that they prefer to use names for variables and functions that are easy to remember:
\redditquote{Favour well thought out names and function}{~- DYS098}
\redditquote{I just make sure to make my variables and functions names not too long and words that are not hard to remember to spell}{~- DYS085}

Finally, some sources mention programming language preferences.
\redditquote{Syntax highlighting and static typing my beloved [...]}{~- PROG041}

\redditquote{I don't think Python is a good language for dyslexics. as it's too forgiving}{~- CS019}
\interviewquote{I have been working a lot with Python. Sometimes indentation is challenging for me, given that indentation is very important in Python. I occasionally get confused with it.}{~- I3}

\subsection{Cognitive Strengths}
Our findings show that even though dyslexic SE professionals face various challenges, they also exhibit various cognitive strengths that are useful in the SE domain. Here, we discuss some of the commonly reported cognitive strengths that are valuable in SE. 


 

Many dyslexic professionals mention that they have an ability to think differently, which helps them solve complex problems. In this context, they compare themselves explicitly to their co-workers.

\blogquote{dyslexic software engineers tend to think outside the box, with the ability to see connections and patterns that others may miss. This contributes to unique and efficient coding solutions.}{~- Blog 2}
%
%
\interviewquote{I'm able to solve problems that others can't think about and think about more logical ways to solve software problems.}{~- I6}
%
%
%
%
\redditquote{a mindset totally different from your co-workers gives you the possibility to see the code and the architecture from another point of you.}{~- PROG061}
%
%
%
%
%

Interestingly, several dyslexic SE professionals mention that they have the ability to analyze information visually:
\interviewquote{I do try to illustrate a lot of things visually, even if it's organizational structures and processes}{~- I4}
\redditquote{ I am a very visual person and these things came easy to me}{~- PROG046}
\interviewquote{visualization helps quite a lot. I draw a lot on stickers for myself}{~- I9}
\blogquote{You're able to visually memorise a lot.}{~- Blog 1}

In relation to SE, individuals mention that this ability visualize information allows them to solve complex software problems easily:
%
%
%
%
%
%


\redditquote{[I] started to see programs in my head as images [...] the whole world, systems of people, tasks, hardware, electrical, everything becomes diagrams to me [like] little pictures.}{~- PROG060}
\redditquote{I think in patterns. Software is patterns. I see the big picture easily. Large software projects require you to understand the big picture.}{~- DYS018}
\blogquote{unique workings of my brain enable me to vividly visualize complex systems internally, facilitating the seamless translation of these mental blueprints into code.}{~- Blog 2}
%
%
%
%
\interviewquote{I think about software almost as like a tree starting with your your main or your run loop. You denote your main at main function, and then everything else branches out from that, and you've got different nodes that have their own sub-trees [...] it is possible to think about code in  3 dimensions, visualizing it to identify issues in areas.}{~- I6}



In addition to having visual thinking ability, we found that many dyslexic SE professionals are creative. This not only makes them good at their job but also helps them come up with unique solutions:
\interviewquote{the creative phase, that is, when people with dylexia excel.}{~- I5}
\redditquote{Creativity certainly helps and it's good to have a creative goal to work towards. That's how I was able to get my first program out.}{~- DYS020}
\redditquote{I'm pretty good at being creative and figuring things out.}{~- PROG061}
\blogquote{Dyslexic individuals often exhibit strong creative problem-solving skills. They can approach challenges from unconventional angles, leading to innovative solutions.}{~- Blog 2}

\section{Discussion}
In the following, we synthesize our findings and answer our RQ.
We then discuss their relation to existing work on SE, including broader work on neurodiversity in SE.
Finally, we provide implications for research and practice.

\subsection{Synthesis}
In our study, we aimed to answer the research question \textbf{RQ: What are the experiences of dyslexic individuals in Software Engineering?}
We started from an initial and intuitive core category called \emph{Reading and Writing Difficulties} that contains evidence on the diverse challenges faced by dyslexic software engineers in reading and writing.
In particular, it highlighted the challenges associated with learning how to program.
To some extent, these learning challenges could be attributed to the textual nature or many traditional ways of learning programming, e.g., through text books, online tutorials or existing documentation.
However, it also reveals interesting challenges for future work.
For example, several individuals reported that reading code that others had written was challenging.
Hence, understanding what makes code more readable to dyslexic programmers and finding better ways of documenting or structuring code in a dyslexic-friendly manner is a promising research direction.

In addition to the reported difficulties, we also shared accounts of many individuals that successfully achieved programming mastery.
This serves as an encouraging testament that working in SE is possible for dyslexic people, despite the apparent mismatch between reading/writing difficulties and programming.
This finding also aligns well with the existing eye-tracking studies by McChesney and Bond~\cite{mcchesney18,mcchesney20,mcchesney2018b,mcchesney21,mcchesney21b}, in particular with the result that dyslexic programmers do not achieve significantly lower code comprehension scores.

Just like in other industries and aspects of daily life, our sources report various generic support tools and strategies.
However, everyday tools such as spell-checkers and AI support tools can make a substantial difference to dyslexic individuals that ultimately enables them to have career success.
Additionally, some of these tools support the individuals in unexpected ways.
For example, electronic communication platforms such as Slack ``encourages people to keep things short'', thus moderating existing reading/writing difficulties.
This finding highlights that inherent limitations of such tools, i.e., the inability to communicate as fluently as in spoken language, turns out to be advantageous to dyslexic individuals.

Beyond the generic support tools, SE-specific tools and strategies are a cornerstone of our studied population.
Tools that are used on everyday basis by software engineers appear especially essential for dyslexic engineers.
For instance, code completion, linters and compilers are consistently reported as life-saving tools in programming environments.
Effectively, these tools become some of the most successful \emph{accommodations}, without which programming would likely be much more affected by reading/writing difficulties of individuals.

The importance of visualization techniques during programming is clearly demonstrated in our data: individuals express a clear preference for visual approaches to understanding logic and code structure, e.g., by using UML diagrams, flowcharts and mind maps. This preference for visual methods aligns with the cognitive abilities often observed in individuals with dyslexia.

Interestingly, our sources report a preference for statically-typed programming languages, as they report typing-related errors sooner than dynamically-typed languages.
Multiple participants and Reddit contributors reported difficulties with Python, which may be related to the language's particular features, such as indentation-based syntax.
One post further described Python as ``too forgiving''.
In the same direction, our data shows that several programming languages are perceived as producing cryptic error messages, e.g., Matlab.
Finally, our findings show that naming variables and functions in certain ways supports understanding and reduces spelling errors.
Overall, these findings emphasize that existing tools and conventions in programming are having a substantial positive impact on dyslexic programmers.
At the same time, they also encourage to re-think and re-visit existing conventions and research with respect to dyslexia.
For example, choice of programming language and preferred naming conventions might differ between dyslexic and non-dyslexic programmers.
As such, existing research might not serve the dyslexic population.

As a final category, we find that dyslexic software engineers exhibit various cognitive strengths that are highly central to SE activities.
These relate to divergent thinking, creativity and visual thinking.
These strengths have been reported in existing dyslexia research, but without a focus on SE.
Our findings thus underscore that these strengths are indeed valuable to the SE context, e.g., for software architecture, debugging, or solving complex SE programs.
However, we also note that the strengths were most commonly reported in the three blogs we studied.
In contrast to our other sources, blogs are more likely to sensationalize findings (e.g., going with the controversial trend of framing aspects of neurodiversity as ``super powers'' \cite{superp1,superp2}) and be less reliable forms of evidence.
As such, we caution that strengths of dyslexic individuals in SE need to be studied in further detail.

A final observation relates to the use of AI and LLMs.
As discussed above, individuals use existing AI support tools, e.g., for rephrasing text.
However, we note a surprising lack of evidence on using AI for SE-specific tasks beyond programming.
That is, more advanced tools might not yet have reached this particular population\footnote{Our data was collected between mid-2024 and early 2025, at a time when these tools were already sufficiently mature.
Therefore, the absence of generative AI use is not readily explained by the study timing.}.

As discussed in Section~\ref{sec:rw}, there is a lack of empirical studies on dyslexia in SE, with the exception of the studies by McChesney and Bond~\cite{mcchesney18,mcchesney2018b,mcchesney20,mcchesney21,mcchesney21b}.
However, dyslexia often co-occurs with other conditions included in neurodiversity, especially with ASD and ADHD \cite{dsm-5}.
In consequence, findings in existing work on this topic might overlap with ours.
Therefore, we discuss how our results relate to broader work on neurodiversity in SE below.

Morris et al.'s study focused on challenges of neurodivergent software engineers in general \cite{morris15}.
However, their interviewees included no dyslexic individual and the follow-up survey only 16 individuals (or 1.9\% of the sample).
Therefore, the challenges reported in their study are more commonly associated with ASD or ADHD, e.g., doing repetitive or mundane coding tasks or managing face-to-face communication.
Additionally, the neurodivergent survey participants reported more comfort with text-based communication than the neurotypical participants, which is unlikely for our dyslexic individuals.
The qualitative studies conducted by Gama et al.~\cite{gama25_gt_asd_adhd} and Newman et al.~\cite{newman2025get} present various challenges and strengths reported by individuals with ADHD and ASD (or both), as well as their relation.
As with Morris et al.'s study, we find almost no overlap related to reported challenges, indicating that the challenges experienced by dyslexic individuals in SE-related activities differ from those with ASD and ADHD.
However, we acknowledge that our findings focus on SE-related tasks, while the related work includes also general workplace challenges, such as attending meetings.
Thus, extending our theory in the future might lead to a larger overlap in these areas.

For all studies, we note an overlap in reported strengths. That is, in \cite{morris15}, some interviewees reported that they excel at out-of-the-box thinking and complex problem solving. Similarly, \cite{gama25_gt_asd_adhd} and \cite{liebel2023challenges} report  creativity, out-of-the-box thinking and complex problem solving.
Finally, Newman et al.~\cite{newman2025get} report creativity as a strength.
Visual thinking abilities are not reported in either of the aforementioned studies, indicating that this might be a particular strength of dyslexic individuals.

Compared to related work on dyslexia beyond the SE community, we find that Beer et al.~\cite{de2022factors} also report that dyslexic professionals have difficulties ``learning and applying knowledge''.
This corresponds to our finding regarding programming learning.
Thus, the challenge seems to not be limited to programming only, but apply to learning more generally.

\subsection{Implications for Research and Practice}

Our findings have several noteworthy \textbf{implications for future research} on dyslexia in SE. We cover these in the following.

First, SE and educational researcher should \textbf{investigate more suitable ways for dyslexic individuals to learn programming}. Our findings consistently show that major struggles arise while learning to program. In particular, many individuals described learning to program as the main obstacle. While there exists work on specific programming languages, e.g., by making use of visual languages, there seems to exist a large potential in better instruction formats, avoiding traditional text-based formats such as books or textual tutorials.

Second, SE and programming languages researchers should \textbf{re-visit the work on choice of programming languages and naming conventions}. Our findings indicate that existing findings might not apply to dyslexic individuals. Specifically, these individuals might favor programming languages for their clarity in error messages, (improved) ability to detect errors through static analysis, i.e., while typing or compiling, and clear block structures, i.e., brackets instead of indentation.
Additionally, they might use different ways to name variables, functions and other programming constructs for better readability.
While existing work covers all of these aspects, we believe personal preferences of dyslexic individuals might add interesting insights.

Third, in a similar direction, future work should \textbf{investigate what makes code more readable and understandable to dyslexic programmers} as opposed to non-dyslexic programmers.
This could lead to better ways of structuring and commenting code - potentially also for non-dyslexic engineers.

Finally, we note that the \textbf{work on strengths of dyslexic and neurodivergent software engineers is still in its initial stages}.
The evidence provided in our study is not very strong and mainly based on blog articles.
However, we note that many of the reported strengths in our data overlap with those of studies on neurodiversity in SE.
Therefore, further studying to what extent these strengths exist and how they can be fostered has a strong potential to advance the SE field and the employability of neurodivergent individuals.

In addition to the implications to research, our findings motivate several \textbf{implications for practitioners}.

First, practitioners should \textbf{include dyslexic people in teams of neurotypical people}, as they possess cognitive strengths that have the potential to substantially improve SE work. In particular, the strength profiles of dyslexic and neurotypical individuals might complement each other and, therefore, improve teamwork.

Second, practitioners should attempt to \textbf{make software documentation more visual}, as dyslexic engineers spend substantial time reading and understanding documentation.
Notably, this is not related to intellectual limitations, but purely due to text comprehension difficulties. In this context, diagrammatic notations might make a comeback in SE.

Third, practitioners should \textbf{make state-of-the-art accommodation tools available to everyone}. While programming support tools are helpful for everyone, they seem to make an especially significant difference to dyslexic engineers. Additionally, this includes standard support tools for office work, such as text-to-speech software.

Third, companies should \textbf{consider programming language choice}. Dyslexic individuals seem to struggle with particular languages or aspects thereof, e.g., dynamically-typed languages or languages that use indendation for programming blocks.

Finally, depending on the programming language the team uses, it might further be helpful to \textbf{adopt pair programming}, as it might help dyslexic individuals in their struggles, while at the same time enabling them to contribute with their unique strengths.

\section{Conclusion}
In this paper, we contribute to the research on neurodiversity in SE by answering the research question ``RQ: What are the experiences of dyslexic individuals in software engineering?''. For this, we interviewed 10 dyslexic software engineers. Further, to complement these data, we analyzed three blogs in which dyslexic software engineers shared their experiences, as well as relevant posts and associated comments gathered from the social media platform Reddit.

Our findings reveal that the reading and writing difficulties faced by dyslexic individuals also extend to difficulties in learning to program. Further, we found that to overcome these challenges, many dyslexic professionals use generic and programming-related support tools and strategies. Our study also highlights various cognitive strengths of dyslexic professionals that are specifically useful in the SE domain. These findings align well with existing educational and clinical research on dyslexia. Additionally, the findings clearly show that dyslexic individuals can succeed in SE and that existing SE-specific tools such as linters or auto-complete are key in enabling this success. Our current work lays a strong foundation for future work in this domain with an overall aim to support more inclusive practices at workplace. 

In future work, we will extend our initial categories into a full theory, potentially adding further categories that relate to workplace dynamics and mental health issues. While we found evidence of these aspects in our data, we did not have the space to incorporate them into this paper.
Additionally, we plan to follow up with a comparative study focusing on understanding to what extent our categories are unique to dyslexic SE professionals.



\bibliographystyle{ACM-Reference-Format}
\bibliography{refs}
\end{document}